    \documentstyle[NATO,psfig]{crckapb}

\def\et al{{\it et al.}}
\begin{opening}
\title{Observation of Cosmological time dilation using Type Ia Supernovae as clocks}
\subtitle{The Supernova Cosmology Project : III}
\author{G.Goldhaber$^1$}
\author{S. Deustua}
\author{S. Gabi}
\author{D. Groom}
\author{I. Hook}
\author{A. Kim}
\author{M. Kim}
\author{J. Lee}
\author{R. Pain$^2$}
\author{C. Pennypacker}
\author{S. Perlmutter}
\author{I. Small}
\institute{E. O. Lawrence Berkeley National Laboratory \& Center
for Particle Astrophysics, 
University of California, Berkeley}
\author{A. Goobar}
\institute{University of Stockholm}
\author{R. Ellis}
\author{R. McMahon}
\institute{Institute of Astronomy, Cambridge University}
\author{B. Boyle}
\author{P. Bunclark}
\author{D. Carter}
\author{K. Glazebrook$^3$}
\author{M. Irwin}
\institute{Royal Greenwich Observatory}
\author{H. Newberg}
\institute{Fermi National Accelerator Laboratory}
\author{A. V. Filippenko}
\author{T. Matheson}
\institute{University of California, Berkeley}
\author{M. Dopita}
\author{J. Mould}
\institute{MSSSO, Australian National University}
\author{W. Couch}
\institute{University of New South Wales}

\end{opening}

\runningtitle{Cosmological Time Dilation}

\begin{document}

% The \begin{document} command comes after the \end{opening}
% command
.
\footnotetext[1]{Presented by G. Goldhaber, 
e-mail address:{\it gerson@LBL.gov}}
\footnotetext[2]{Current address:  CNRS-IN2P3, University of Paris}
\footnotetext[3]{Current address: Anglo-Australian Observatory}

\begin{abstract}
This work is based on the first results from a systematic search for high
redshift Type Ia supernovae. Using filters in the $R$-band
we discovered seven such SNe, with redshift $z=0.3 - 0.5$, 
before or at maximum
light.  Type Ia SNe
are known to be a homogeneous group of SNe, to first order, 
with very similar light 
curves, spectra and peak luminosities.  In this talk we 
report that the light curves we observe are all broadened (time
 dilated) as expected from the expanding universe hypothesis. 
Small variations from the expected $1+z $ broadening of the light curve 
widths can be attributed to a width-brightness correlation 
that has been observed for nearby SNe ($z<0.1 $). We show 
in this talk the first clear 
observation of the cosmological time dilation for macroscopic objects.

\end{abstract}

%\keywords{supernovae:general---cosmology: distance scale}
%\keywords{globular clusters,peanut clusters,bosons,bozos}

% That's it for the front matter.  On to the main body of the paper.
% We'll only put in tutorial remarks at the beginning of each section
% so you can see entire sections together.
%
% In the first two sections, you should notice the use of the LaTeX \cite
% command to identify citations.  The citations are tied to the
% reference list via symbolic tags.  We have chosen the first three
% characters of the first author's name plus the last two numeral of the
% year of publication.  The corresponding reference has a \bibitem
% command in the reference list below.
%
% Please go to the LaTeX manual for a complete description of the
% \cite-\bibitem mechanism.

%\section{Introduction}

%\vspace{-0.18in}
%A subset of these supernovae can provide the distant
%standard clocks needed to finally perform this 65-year-old test of the nature
%of redshift.
\section{Introduction}
  
In an ongoing systematic search, we have recently discovered and studied
seven supernovae (SNe) at redshifts between $z = 0.35$ and 0.46 as discussed in Perlmutter's talk at this conference. For details on search technique,
light curves, and spectra, see Perlmutter et al. 1996.

\section{Supernova Homogeneity}
As discussed in several talks at this conference, Type Ia supernovae are, as a class, highly homogeneous.  They are explosion events that are
apparently triggered under very similar physical conditions.
Their ``light curves'' 
 scatter by less
then $\sim25$\% RMS in brightness (Vaughan et al 1995a, 1995b), and less than
15\% RMS in full-width-at-half-maximum (Perlmutter 1996), in a sample of 
``normal'' Type Ia supernovae, after rejecting the abnormal $\sim$15\% with
red colors (see Vaughan et al 1995a).  Their spectral signatures also follow 
a well-defined evolution in time.   
A paper giving all the photometric and spectroscopic 
 measurements for our SNe,as well as a detailed discussion of 
the evaluation of $q_0$, is in preparation.

 Most of our SNe have been 
  followed for about a year.  For three of our SNe we have obtained spectra at an early enough time to
observe both the characteristic SN spectrum as well as the host galaxy
spectrum. These three supernova spectra, in their rest system, all have
  spectral features matching nearby
Type Ia's at the same epoch in great detail. 
 In the other cases, we could not obtain the
spectra early enough, due to inclement weather, so that by the time the
spectra were taken the SN light could not be separated from the galaxy 
spectrum.  Redshifts $z$ were obtained from the host galaxy spectra.
 All these
details on the spectra will  be presented in our forthcoming paper (Perlmutter et al. 1996).

In the present paper we will make use of the ``standard" nature of the Type Ia SNe light curves.  This feature allows us to  consider the Type Ia supernovae  as ``clocks" at cosmological distances.

\section{The Observed SNe }

Due to the large redshifts we are aiming for, we have carried out  our
search using $R$-band filters, while most of the
data on nearby SNe was taken in $B$-band or V-band filters. To compare our data with  nearby
 measurements in  $B$-band magnitudes, we use a template compiled by  B.Leibundgut.
To this template curve we have to
apply a $K$ correction  which compares the SN spectrum 
as observed ``nearby" in a blue filter with the red shifted 
spectrum as observed at high $z$ values in a red filter (as 
discussed by A. Kim at this meeting and Kim, Goobar, \& Perlmutter 1996).
  The resulting new templates---one for each redshift---are then 
in the $R$-band in which our data was taken.

 The recent work of Phillips (1993), Hamuy et al (1995), and 
Riess et al.(1995) has emphasized the {\em inhomogeneity}
of Type Ia light curve shapes, particularly for the ``non-normal''
redder Type Ia's.  Perlmutter (1996) provides a single-parameter
characterization of these light curve differences, which is simply a
time-axis stretch factor, $s$,  which stretches (or compresses) the
Leibundgut template light curve. From this study based on {\em nearby} SNe
it was shown that this stretch factor $s$ extends over a range of 0.65 to 1.1. To observe the effect of cosmological time dilation experimentally the expected dilation factor $ 1+z $ is modified by the stretch factor $s$. The observable effect is then $ d = s(1+z) $. Since $s$ is asymmetric around 1, for low values of $s$ which occur for the most
extreme 15\% of non-normal Type Ia supernovae with red colors, 
the $1+z $ effect can be essentially cancelled by $s$, giving an observed
dilation of $d \approx 1$.

\section{Fit to the data.}
We fit each of our observed seven  SNe to the $R$-band template light curve using the fitting 
program MINUIT (James and Roos 1994). Each SN is fit to this template, 
expressed now  in normalized counts,
(rather than in magnitudes). We fit three variables:
the height of the light curve, the day of peak light and a
time dilation, $d$, of the  width of the light curve.

As an illustration of the cosmological time dilation 
effect, we show in Fig 1 one of our seven SNe $R$-band
light curve data points, SN 1994H, plotted against the observed time axis. 
The dashed curve is the best fit Leibundgut template 
with $K$ corrections, with no time dilation, i.e. $d$ fixed at $d = 1$. This gives a  ${\chi}^2/DoF = 3.1  $.
The solid curve is the best fit with $d$ fixed at $d = 1+z$ for a 
${\chi}^2/DoF = 1.3$ . This corresponds to the slowing down of our 
``clock", with the cosmological time dilation expected 
for a redshift of $z = 0.374$ .

\begin{figure}
%\vspace{8cm}
\psfig{figure=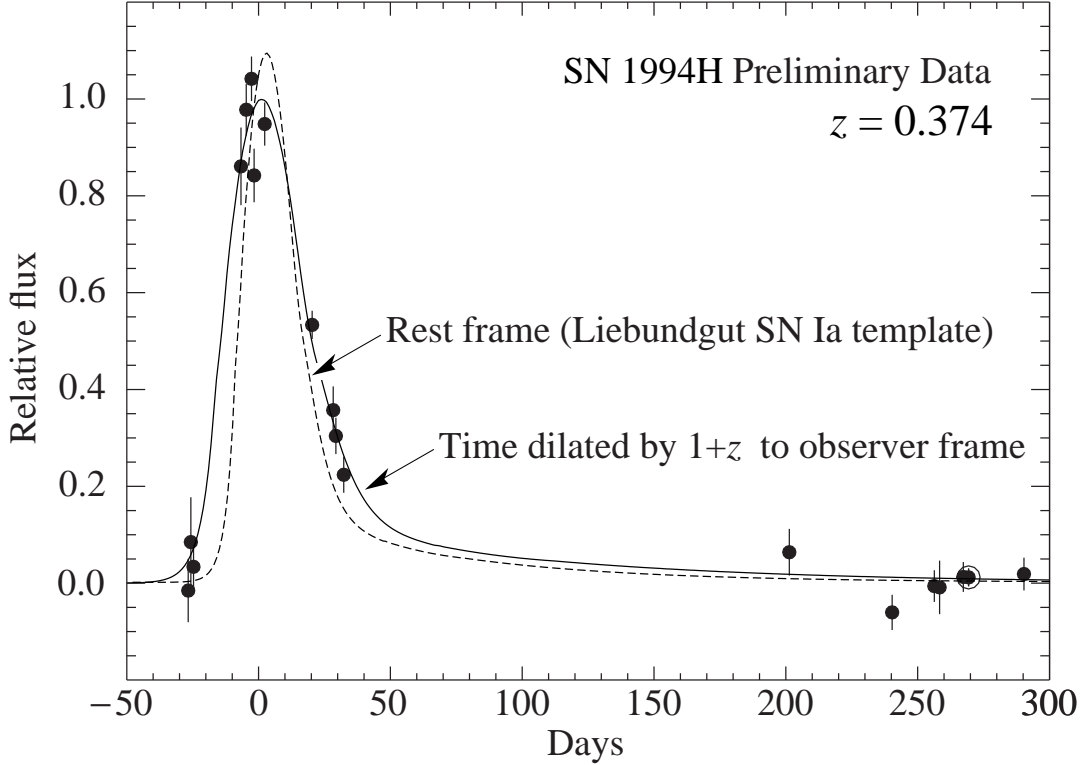,height=4.0in}
\caption{The best fit light curves for $SN94H$, under the hypothesis $d = 1$ dashed curve, and $d = 1+z$, solid curve, where $z = 0.374$. The open circle corresponds to the reference image used.}
\end{figure}
  
%In Table I we present the results of the fits to our seven SNe.
%Col 1 gives the SN number in our numbering system. Col 2 gives the SN ``name".
%col 3 gives the $\chi^2$/DoF allowing all 3 parameters to vary. Col 4 gives
% the resulting stretch value, s. Col 5 gives the $\chi^2$/dof for
%the same fit in which now the stretch factor was kept fixed
%at the value $s = 1+z$ expected for a cosmological time dilation.
%In col 6 we
% present the same fit where now the factor $d$  is kept fixed at 1. This
%corresponds to the case in which there would be no time dilation of the
%template curve for SNe Ia. Here we note a very poor fit to the data .

Given the asymmetric spread of light curve widths discussed above, 
we can predict what the 
distribution should look like at high redshifts for a universe
with and without time dilation.   These predictions are indicated by the
shaded bands of Figure 2.  Note that at redshifts between 
$0.1 < z < 0.5$, any examples
of Type Ia supernovae with an observed 
width greater than $s(1+z) = 1.1$ provide
evidence for the time-dilation model;  the narrower supernovae
neither help nor hurt in distinguishing the models since the
two ranges both are consistent with supernovae
with observed width smaller than 1.1.  At redshifts
higher than $z = 0.5$, all supernovae become useful for separating the
two models, since there is no overlap.
Fig 2 also shows the actual data for the best-measured 
five of our seven SNe, plotted on top of these predicted ranges for the
two models.

We can make this test even more powerful, even at redshifts 0.1 $< z < 0.5$,
by using another {\em independent} observable to indicate the intrinsic width of
the supernova.  Vaughan et al (1995) and Branch et al (1996) have shown
that supernova color can indicate the width (and brightness) of a given
supernova.  $B-V$ color can distinguish among the narrow $s < 1$ supernovae,
while $U-B$  color appears to identify the width and brightness
of a supernova within
the entire range from $0.6 < s < 1.1$.   
(It is possible to confuse  intrinsic color differences with reddening, if 
sufficient photometry points are not available, however this is not a
problem for the brighter, slower supernova light curves that appear {\em bluer},
and thus are not confused with reddened supernovae.)

\begin{figure}
%\vspace{8cm}
\psfig{figure=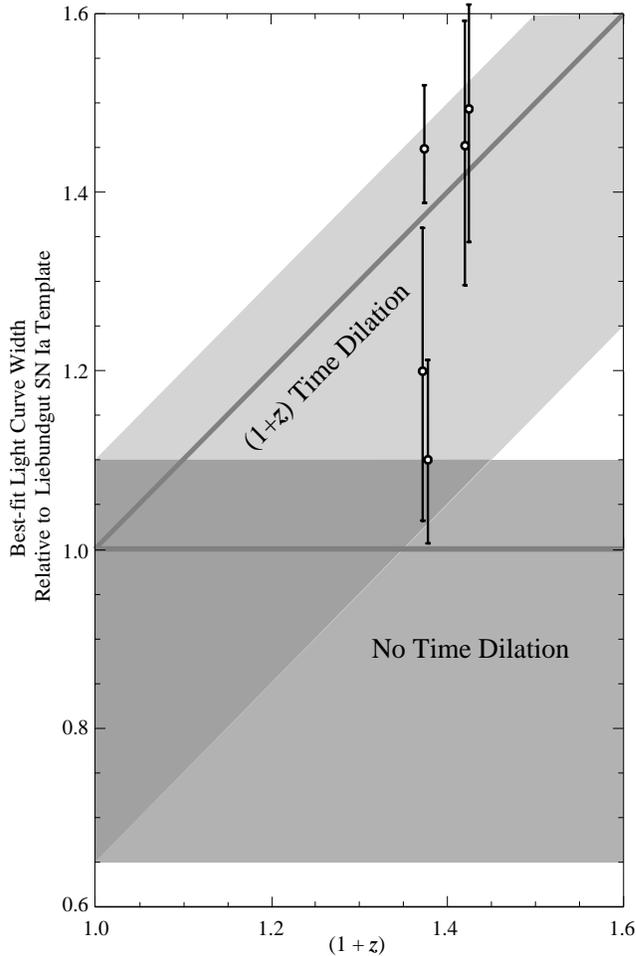,height= 5.0in}
\caption{ 
The observed fitted time dilation $ d $ plotted against $ 1+z $ 
The diagonal band corresponds to the time-dilation hypothesis taking into account the region of width-stretch,$s$, observed for nearby SNe. The band along the x-axis corresponds to the no-time-dilation case. The observed deviations from the time dilation case can be attributed to the known distribution in width caused by the width-peak-magnitude correlation folded into the experimental error.}
\end{figure}

So far we have completed color analysis of points on the light curve for two of
the supernovae, SN94H and SN94G.  SN94H has an observed $B-R$ color 
near peak consistent with a rest-frame $U-B$ color of $\sim$$-0.6$,
indicating that this supernova is like the nearby SN 1991T, which has a stretch
factor of $s = 1.08 \pm 0.05$.  SN94G has an $R-I$ color lightcurve that is
consistent with a supernova with $s = 1.0 \pm 0.1$, and not consistent with
narrower light curve supernovae.  (Note that neither of these light curves
is consistent with significant reddening.)
In Figure 3, we plot these two supernovae after dividing out the
intrinsic width, $s$, deduced from their color, and adding in quadrature the extra
error bar's uncertainty.  On this plot the two models now appear as 
lines, not ranges.

It is clear from Figures 2 and 3, that the time dilation model is 
the much better fit to the data. The ${\chi}^2/DoF = 0.13$ for the time dilation hypothesis, the diagonal line
on Figure 3, while ${\chi}^2/DoF = 12.6 $ for the no-time-dilation hypothesis,
the line along the x-axis.
 Thus for the majority of our distant SNe the values for $d$ are dominated by the cosmological time dilation. 
  We are obtaining  color information on all the more recent supernovae to
make it possible to plot future observations on Figure 3, after
identifying the intrinsic width from the color.

\begin{figure}
%\vspace{8cm}
\psfig{figure=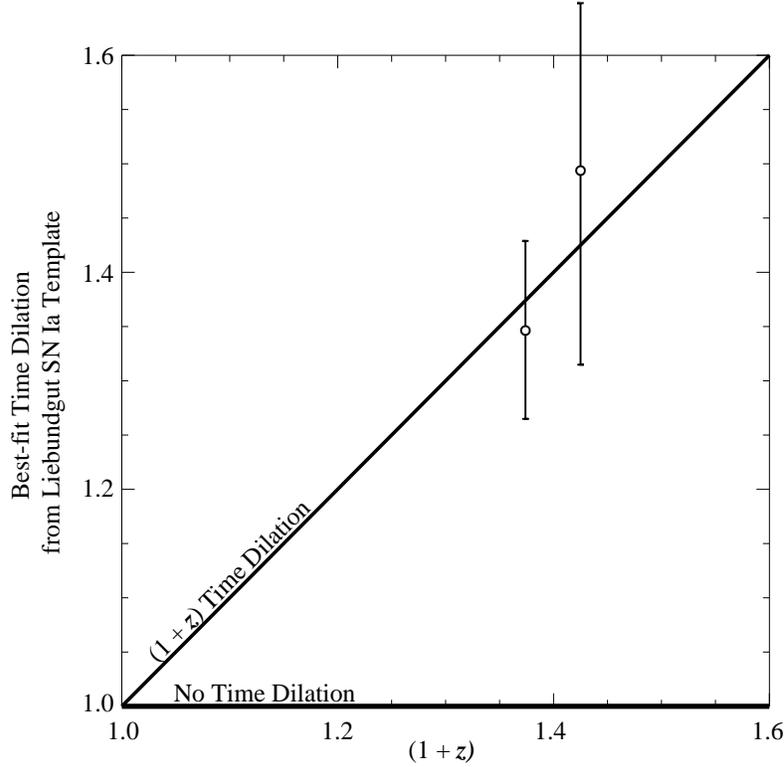,height= 4.0in}
\caption{A plot of $d/s$  for two of our SNe, SN94H and SN94G for which the stretch, $s$, was deduced from the SNe colors. } 
\end{figure}

\section{Historical note.}

A measurement such as we have performed was first proposed by 
 O. C. Wilson, in 1939, as a simple test of the conjecture that 
astronomical redshifts are explained by an expanding universe model, rather
than, e.g., ``the gradual dissipation of photonic energy,'' later called the
``tired light'' model. The expanding universe has by now won almost
universal acceptance, this classical test,
while attempted by Rust ( 1974 ) has however
never been demonstrated until now, due to the lack of sufficiently distant ``standard clocks.''  
%Wilson's test suggested a distant astronomical event of known duration ,in %particular a SN as 
%a ``standard clock'' and looks for the time dilation of this duration
%at a given redshift.  
In an 
expanding universe, the time dilation, $d$ , should match the 
redshift of spectral features, $d = 1+z$, while a tired light model
implies no time dilation, $d = 1$, at any redshift.
(Even in an expanding universe, some ``tiring'' of light could occur, and this
test could also be turned around to bound the extent to which photons lose
energy traveling through intergalactic space.)  
While some of the errors in our fit of the lightcurve dilation-factor are large, it is clear that we have observed the time dilation of macroscopic clocks at cosmological  distances.

\section{Discussion}
There is one  source of concern  that must still be addressed:
how certain are the Type Ia identifications?   
We expect to find primarily Type Ia since these are the brightest SNe by typically 2 magnitudes.
For three of our SNe we have  spectral identification
and their light curves are consistent with the other four. One
of the remaining four SNe was discovered in an eliptical galaxy, and
therefore is highly likely to be a Type Ia.  In sum, the current set of
data are very likely to all be SNe Ia.

\acknowledgements
This work was supported in part by the National Science Foundation
(ADT-88909616) and the U.S. Dept. of Energy (DE-AC03-76SF000098).

\clearpage

%\begin{figure}
%\plotone{sgi9259.ps}
%\caption{We use the \LaTeX\ {\tt figure} environment syntax to set this
%figure caption.}
%\label{relation}
%\end{figure}

% That's all, folks.
%
% The technique of segregating major semantic components of the document
% within "environments" is a very good one, but you as an author have to
% come up with a way of making sure each \begin{whatzit} has a corresponding
% \end{whatzit}.  If you miss one, LaTeX will probably complain a great
% deal during the composition of the document.  Occasionally, you get away
% with it right up to the \end{document}, in which case, you will see
% "\begin{whatzit} ended by \end{document}".

\end{document}